\documentclass[11pt]{article}
\usepackage{moriond,epsfig}
\def\as{\alpha_S}

\begin{document}
\vspace*{4cm}
\title{Progress in top-pair production at hadron colliders \footnote{Preprint numbers: CERN-PH-TH/2012-155,  TTK-12-24.}}

\author{Peter B\"arnreuther, Micha\l{}  Czakon}
\address{Institut f\"ur Theoretische Teilchenphysik und Kosmologie,
RWTH Aachen University, D-52056 Aachen, Germany}
\author{Alexander Mitov \footnote{Speaker.}}
\address{Theory Division, CERN, CH-1211 Geneva 23, Switzerland}

\maketitle\abstracts{
We review recently derived NNLO QCD results for top-pair production at hadron colliders. We compare the size of the NNLO corrections in $q\bar q\to t\bar t+X$ with the LO and NLO ones, and address the question of convergence of perturbative series. We compare the NLO and NNLO K-factors for the Tevatron.}

\section{Introduction}

The speed with which the LHC, complemented by the measurements from the Tevatron, is reshaping the landscape of particle physics is remarkable. In two short years a number of constraints on new physics have been placed and the search for a light Standard Model Higgs boson has narrowed down to a small window around $m_{\rm Higgs}=126$ GeV. An important lesson from these searches is the need for precise knowledge of Standard Model signals and backgrounds.

During the last several years, similarly profound advances in our ability to tame perturbative QCD and describe with increasing precision and confidence hadron collider observables were made. These developments are nowhere more evident than in top physics.

First, the advances in NLO calculations of the recent past \cite{Bern:1994zx,Britto:2004nc,Ossola:2006us,Giele:2008ve}
made possible fully exclusive calculations for final states with large multiplicities ($t\bar t$ + up to 2 jets), including top decays and even accounting for non-factorizable effects \cite{Dittmaier:2007wz,Bredenstein:2009aj,Melnikov:2009dn,Bevilacqua:2009zn,Bredenstein:2010rs,Bevilacqua:2010ve,Denner:2010jp,Bevilacqua:2010qb,Bevilacqua:2011aa,Worek:2011rd}. Only few years ago such progress seemed like an impossible task.

Second, also in the last several years, a renewed, massive push for describing the higher order (i.e. NNLO) corrections in top pair production was undertaken. It is an approach based on approximating the NNLO cross-section with its threshold behavior \cite{Beneke:2009rj,Czakon:2009zw,Beneke:2009ye}. A number of phenomenological predictions were made \cite{arXiv:0906.5273,Ahrens:2010zv,Beneke:2010fm,Ahrens:2011mw,Ahrens:2011px,Kidonakis:2011ca,Beneke:2011mq,Cacciari:2011hy,Beneke:2011ys}, and compared in \cite{Kidonakis:2011ca,Cacciari:2011hy,Beneke:2011ys}. We have concluded \cite{Cacciari:2011hy} that the ability of such an approach to unambiguously solve the question of higher order (NNLO) effects in top-pair production is limited. To resolve this, we have undertaken the task of computing the complete NNLO result for top-pair production at hadron colliders. The result for the $q\bar q\to t\bar t+X$ reaction appeared in \cite{Baernreuther:2012ws}, which we describe next. Very recently, also the so-called BLM/PMC approach was applied to top production at NNLO \cite{Brodsky:2012sz}.

\section{Results at  NNLO}

In \cite{Baernreuther:2012ws} we calculate the NNLO corrections to the reaction $q\bar q\to t\bar t+X$. The calculation is based on the two-loop virtual corrections from \cite{Czakon:2008zk}, the analytical form for the poles \cite{Ferroglia:2009ii} and the one-loop squared amplitude \cite{Korner:2008bn}. The real-virtual corrections are derived by integrating the one--loop amplitude with a counter-term that regulates it in all singular limits \cite{Bern:1999ry}. The finite part of the one-loop amplitude is computed with a code used in the calculation of $pp\to t\bar t + {\rm jet}$ at NLO \cite{Dittmaier:2007wz}. The most nontrivial part of the calculation are the double real corrections \cite{Czakon:2010td}. As in Ref.~\cite{Czakon:2010td}, we do not include the contribution from the reaction $q\bar q \to t\bar t q\bar q$ where the final state light pair has the same flavor as the initial state one. We expect the contribution from this reaction to be negligible. The explicit results for this contribution, as well as the remaining pure fermionic reactions $qq,qq'$ and $q\bar q'$, with $q'\neq q$, will be presented elsewhere.

The dominant role of the $q\bar q\to t\bar t+X$ reaction to top-pair production at the Tevatron, makes it possible to use the results derived in \cite{Baernreuther:2012ws} for a consistent, NNLO-level phenomenology at this collider of the hadronic total inclusive cross-section:
\begin{equation}
\sigma_{\rm had}(\rho_h) = \int_0^{\beta_{\rm max}} d\beta\, \hat\sigma(\beta)\, \Phi_{\rho_h}(\beta)\, ,
\label{eq:sigmahad}
\end{equation}
where $\rho_h \equiv 4m_{\rm top}^2/s_{\rm collider}$ and $\beta_{\rm max} \equiv \sqrt{1-\rho_h}$. The flux $\Phi$ reads:
\begin{equation}
\Phi_{\rho_h}(\beta) = {2\beta \over 1-\beta^2}~ {\cal L}\left({\rho_h\over 1-\beta^2}\right) \, ,
\label{eq:flux}
\end{equation}
and ${\cal L}(x) = x \left( f_1\otimes f_2 \right) (x)$ is the usual partonic luminosity given as a convolution of two parton distributions. We have suppressed for short the partonic indices (and the sum over them) as well as factorization and renormalization scales (we set $\mu_F=\mu_R=m_{\rm top}$ throughout).

Through NNLO the partonic cross-section $\hat\sigma$ has the following expansion:
\begin{eqnarray}
\hat\sigma(\beta) &=& {\as^2\over m^2}\left( \sigma^{(0)} + \as \sigma^{(1)} + \as^2 \sigma^{(2)} + \dots\right)  \equiv {\as^2\over m^2}\left( f_{\as^2} + f_{\as^3} + f_{\as^4} + \dots \right) \, .
\label{eq:sigma}
\end{eqnarray}

\begin{figure}
\psfig{figure=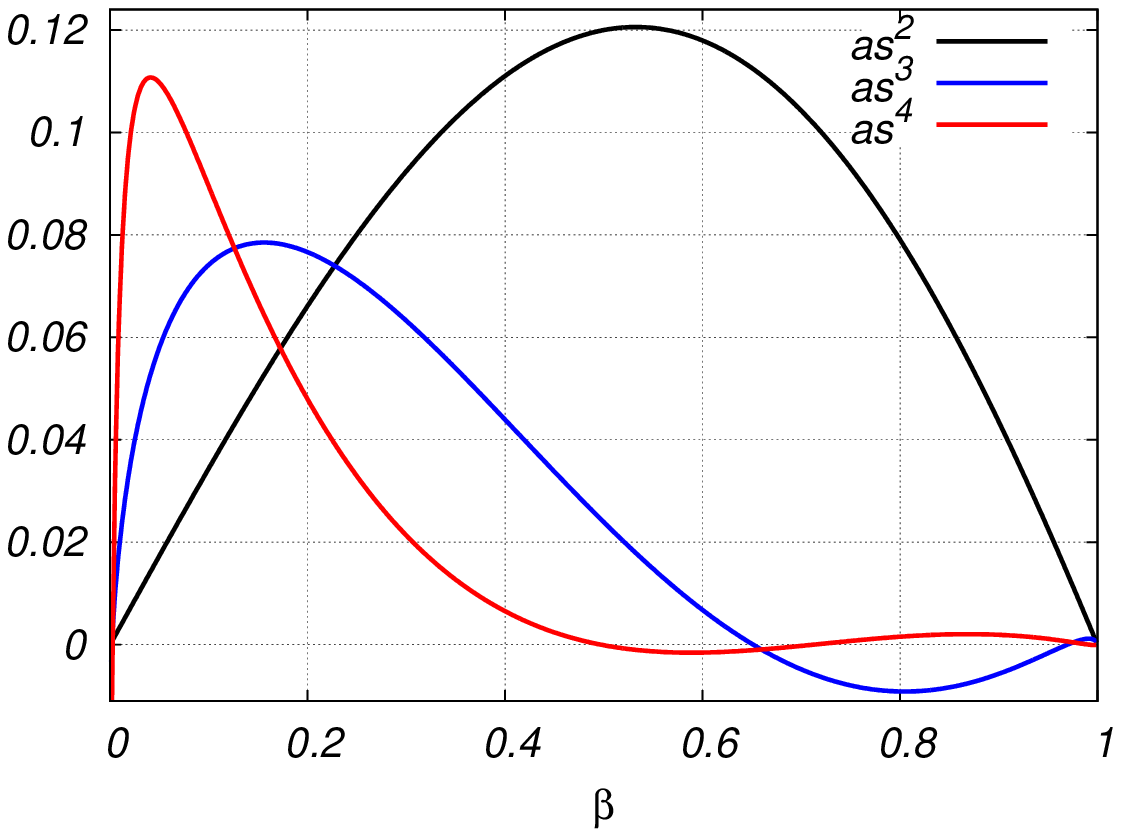,height=2.25in}
\psfig{figure=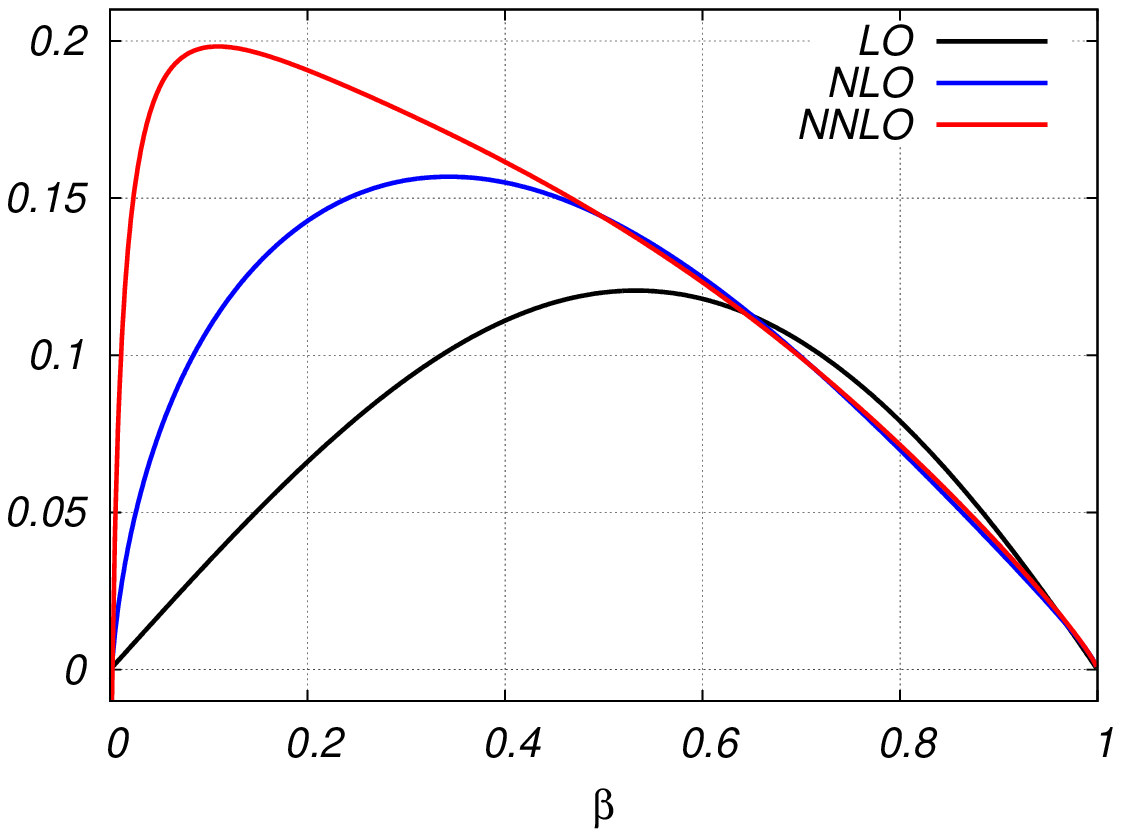,height=2.25in}
\caption{(Left plot) the functions $f_{\as^2}$ (black), $f_{\as^3}$ (blue) and $f_{\as^4}$ (red) as defined in Eq.~(\ref{eq:sigma}).
\hskip 30mm (Right plot) their contribution to $\hat\sigma$ at LO ($f_{\as^2}$), NLO ($f_{\as^2} + f_{\as^3}$) and NNLO ($f_{\as^2} + f_{\as^3} + f_{\as^4}$).
\label{fig:fig}}
\end{figure}

The functions $ f_{\as^2}, f_{\as^3}$ and $f_{\as^4}$ for the reaction $q\bar q\to t\bar t+X$ and with $N_L=5$ are plotted on the left Fig.~(\ref{fig:fig}), while on the right we present their contribution to the cross-section, i.e. $f_{\as^2}$ at LO, $f_{\as^2} + f_{\as^3}$ at NLO and $f_{\as^2} + f_{\as^3} + f_{\as^4}$ at NNLO.

The curves plotted on Fig.~(\ref{fig:fig}) naturally raise the question about the convergence of the perturbative series. We observe that the subsequent higher orders do not get smaller (for example in the naive sense of their integrals or maxima) and tend to get distributed closer and closer to the absolute threshold $\beta = 0$. The reason for this behavior is that the higher orders are more and more dominated by the soft gluon and Coulomb terms. One can anticipate that at even higher perturbative orders this trend will continue. 

We would like to point out two important additions to this observation. First, it does not account for the fact that close to threshold (where the corrections are largest) perturbation theory breaks down and resummation of the soft-gluon corrections is needed to restore the predictivity of perturbation theory. A not-so-well-known example for an effect of this type is the fact that beyond order $\as^4$, the Coulomb terms in the fixed order expansion will render the perturbative cross-section nonintegrable - a seemingly disastrous implication - that is resolved by the observation that Coulombic terms {\it must} be factorized and resummed; see \cite{Beneke:2011mq} for details.

Second, the size of the curves on Fig.~(\ref{fig:fig}) does not directly determine the hadronic cross-section. For that one needs to multiply them with the partonic flux (\ref{eq:flux}). Due to Jacobian factor $\sim \beta$ the flux vanishes at threshold as a power which dominates over the logarithmic rise of the partonic cross-section due to soft-gluon radiation. For example, for top production at Tevatron, the flux is roughly bell shaped with maximum around $\beta\approx 0.7$ thus making the behavior of the partonic cross-section away from threshold more relevant for the hadronic cross-section. 

Finally, we discuss K-factors at the Tevatron. Introducing the shorthand notation $I_n$ for the contribution of the functions $f_{\as^n},~ n=2,3,4$ (from all partonic reactions) to the cross-section $\sigma_{\rm had}$, and following the setup outlined in \cite{Baernreuther:2012ws} implemented in the program {\tt Top++} (ver. 1.2) \cite{Czakon:2011xx}, we get:
\begin{equation}
I_2 = 5.221 \, [{\rm pb}]~~,~~I_3 = 1.234 \, [{\rm pb}]~~,~~I_4 = 0.548 \, [{\rm pb}] \, .
\end{equation}
From the above equation we derive the following K-factors for the Tevatron:
\begin{equation}
K_{\rm NLO/LO} = 1.24~~,~~ K_{\rm NNLO/NLO} = 1.08 \, .
\end{equation}
We use the MSTW2008nnlo68cl pdf set \cite{Martin:2009iq}. We note that the K-factors are not sensitive to the choice of pdf set (for example NNLO or NLO), or if NNLL resummation is included or not.

\section{Conclusions and outlook}

In \cite{Baernreuther:2012ws} we have performed the first ever NNLO calculation of a hadron collider process with more than 2 colored partons (there are four) and/or massive fermions. The result exhibits remarkable stability with respect to scale variation and suggests very precise estimate of the total cross-section for top-pair production at the Tevatron. When supplemented with soft gluon resummation at NNLL the stability of the result further increases, as expected. The result calculated matches all partial checks available in the literature.

The K-factor derived from the NNLO result is modest and shifts the NLO result by about 8\% (when both the NLO and NNLO are computed with the same pdf). The inclusion of the $q\bar q\to t\bar t+X$ reaction at NNLO improves also the prediction at the LHC \cite{Czakon:2011xx}. While this is the most complete theoretical prediction available for the LHC, it is clear that substantial improvement for the Large Hadron Collider can be expected only upon inclusion of the $qg-$ and $gg-$initiated reactions. Our calculational method is well suited for the calculation of these reactions and, even more importantly, fully differential top-pair production, including the ${\cal O}(\as^4)$ correction to the top-pair charge asymmetry. We anticipate reporting these results in the near future.

\section*{Acknowledgments}
The work of M.C. was supported by the Heisenberg and by the Gottfried Wilhelm Leibniz programmes of the Deutsche Forschungsgemeinschaft, and by the DFG Sonderforschungsbereich/Transregio 9 ÒComputergest\"utzte Theoretische TeilchenphysikÓ.

\end{document}